# Towards the Complete Relational Graph of Fundamental Circuit Elements[*]

SHANG Da-Shan(尚大山)[**], CHAI Yi-Sheng(柴一晟)[**], CAO Ze-Xian(曹则贤), LU Jun(陆俊), SUN Young(孙阳)[***]

*Beijing National Laboratory for Condensed Matter Physics, Institute of Physics, Chinese Academy of Sciences, Beijing 100190, China*

**Abstract:** A complete and harmonized fundamental circuit relational graph with four linear and four memory elements is constructed based on newly defined elements, which provides a guide to developing novel circuit functionalities in the future. In addition to resistor, capacitor and inductor which are defined in terms of a linear relationship between the charge $q$, the current $i$, the voltage $v$, and the magnetic flux $\varphi$, Chua proposed in 1971 the fourth linear circuit element to directly relate $\varphi$ and $q$. A non-linear resistive device defined in memory $i$-$v$ relation and dubbed memristor, was later attributed to such an element and has been realized in various material structures. Here we clarify that the memristor is not the true fourth fundamental circuit element but the memory extension to the concept of resistor, in analogy to the extension of memcapacitor to capacitor and meminductor to inductor. Instead, a two-terminal device employing the linear magnetoelectric effects, termed transtor, possesses the function of relating directly $\varphi$ and $q$ and should take the position of the fourth linear element. Moreover, its memory extension, termed memtranstor, is proposed and analyzed here. .

[*]Supported by the National Natural Science Foundation of China under Grant No 11227405, 11374347, 11274363 and 11474335, and the Strategic Priority Research Program of the Chinese Academy of Sciences under Grant No XDB07030200.

[**]These authors contribute equally to this work.
[***]Corresponding author. Email: youngsun@iphy.ac.cn.

# 1. Introduction

## 1.1 History of memristor

Following circuit theory, three fundamental two-terminal linear circuit elements, *i.e.*, the capacitor $C$, the inductor $L$, and the resistor $R$, are defined in terms of a linear relationship between two of the four basic circuit variables, namely, the charge $q$, the current $i$, the voltage $v$, and the magnetic flux $\varphi$, as schematically illustrated in Fig. 1(a). In 1971, Chua pointed out that there should be a fourth fundamental linear element which could be defined in terms of $\varphi$ and $q$,

$$\mathrm{d}\varphi = M\mathrm{d}q. \qquad (1)$$

where the coefficient $M$ is termed as memristance.[1] As no real circuit element that directly connects $\varphi$ and $q$ could be identified at that time, Chua made a transformation of Eq. (1) by using the relationships

$$v = \mathrm{d}\varphi/\mathrm{d}t \text{ (Faraday's law)} \qquad (2)$$

$$i = \mathrm{d}q/\mathrm{d}t \qquad (3)$$

and obtained

$$v = Mi. \qquad (4)$$

However, Chua found that the quantity $M$ in Eq. (4) is equivalent to the resistance $R$, and thus of no particular interest. In order to introduce some non-trivial meanings to $M$, Chua speculated that $M$ might be a function of $q$ and $t$ instead of being a constant. In this case, a nonlinear resistive element, dubbed memristor, can be defined from Ohm's law

$$v(t) = M(q(t))i(t) . \qquad (5)$$

Later, Chua and Kang further generalized the concept of memristor to a much broader class of nonlinear dynamic systems, the memristive systems,[2] which are described by

$$v(t) = M(x, i, t)i(t) \qquad (6)$$

$$dx/dt = f(x, i, t) \tag{7}$$

where *x* is vector representing the state variables of the system, and *f* is a continuous vector function.

Chua's work lies dormant for nearly 40 years without causing much attention. In 2008, Strukov *et al*. reinvented the memristor in a Pt/TiO$_{2-x}$/Pt sandwiched structure which exhibits a hysteretic *i-v* curve.[3,4] This discovery incurred strong interest in inquiring the resistance switching mechanism which had been the main roadblock to the commercialization of resistive random access memory devices.[5-8] Furthermore, Chua argued that all the two-terminal nonvolatile memory devices based on resistance switching are memristors, regardless of the materials applied or physical mechanisms in operation, and postulated that the hysteretic *i-v* characteristics pinched at origin be the fingerprint of a memristor.[9] Inspired by the discovery of memristor, the nonlinear memory counterparts of capacitor and inductor, now known as memcapacitor and meminductor accordingly, have also been proposed and experimentally demonstrated.[10-12]

## 1.2 Is the memristor the true fourth circuit element?

The memristor has great potential for developing new functional devices to be applied in synapse memory, neuromorphic computing, Boolean logic operations, and so on.[13-15] However, its physical identity as the fourth fundamental circuit element remains questionable. Firstly, the memristor is obtained in practice from the *i-v* relationship rather than, *as originally proposed*, from a direct relationship between magnetic flux $\varphi$ and charge *q*. In other words, a memristor in *i-v* definition can work properly without invoking the basic variable $\varphi$. Secondly, the *i-v* definition of memristor from the Ohm's law (Eq. (5)) is physically **NOT** equivalent to that of the $\varphi$-*q* in Eq. (1) because the Faraday's law (Eq. (2)) to convert $\varphi$ to *v* is only applicable in devices with closed area under changing magnetic field (like inductor) or moving part under magnetic field (like generator) while $v$ to $\varphi$ when electric field $E$ has non-zero curl ($\nabla \times E$). In this sense, the memristors studied so far actually do not allow a $\varphi$-$q$ relationship in that it is incorrect to convert $v$ back to $\varphi$ for a resistor. Thirdly, the linear magnetoelectric (ME) effects, which refer to the induction of electric polarization ($P$) by the application of a magnetic field ($H$) or magnetization ($M$) by an electric field ($E$) linearly, naturally lead to a linear coupling between $M$ and $P$. Therefore, the linear ME effect may consequently give a direct and linear link between $\varphi$ and $q$, should be used to construct the fourth linear circuit element, as already speculated by Mathur.[16]

Based on the above arguments, the memristor that operates upon the nonlinear memory $i$-$v$ relationship should take a position of its own as the memory counterpart of resistor, in analogy to the memcapacitor with regard to capacitor, and the meminductor to inductor, as illustrated in Fig. 1(b). It is then very natural to be aware of that the fourth *memory* circuit element which memorizes past states through which the systems have experienced in the correlation between $\varphi$ and $q$ is still lacking. And the new memory element should be the counterpart of the fourth linear element based on some specific non-linear ME effects.

In this paper, we conceive a two-terminal model device made of a ME media showing either the linear or butterfly-shaped memory ME effects and prove that its circuit functions fully satisfy the requirements as the fourth fundamental elements, linear and memory, respectively. We coined the term transtor, $T$, and memtranstor, $M_T$, to denote respectively the linear and the memory actualizations of these new elements. With the introduction of the concepts of transtor and memtranstor, a complete and harmonized relational graph for the fundamental circuit variables with four linear and four memory elements, can be obtained, as shown in Fig. 1(b).

## 2. Derivation of linear $\varphi$-$q$ relationship based on linear magnetoelectric effect

The possibility of the ME effect was first predicted by Curie in 1894 on the basis of symmetry considerations,[17] and the term "magnetoelectric" was coined by Debye in 1926.[18] Since the renaissance of multiferroic materials research in 2004, ME effect has drawn great interest due to its promise for many applications.[19,20] Here, we consider a device made of a block of ME medium sandwiched between two parallel electrodes, to operate with either the longitudinal (Fig. 2(a)) or transverse (Fig. 2(b)) ME effect. For a non-ferroic, linear ME medium, according to Landau's theory, the Landau free energy $F(E,H)$ is given by:[20]

$$-F(E,H) = \frac{1}{2}\varepsilon_0\chi_e E^2 + \frac{1}{2}\mu_0\chi_v H^2 + \alpha EH, \tag{8}$$

where $\varepsilon_0$ ($\mu_0$) and $\chi_e$ ($\chi_v$) are the permittivity (permeability) of vacuum and electric susceptibility (magnetic susceptibility) of the medium, respectively, α is the linear ME coefficient. By differentiating $F(E,H)$ with respect to $E$ and $H$, one obtains:

$$-\frac{dF}{dE} = P = \varepsilon_0\chi_e E + \alpha H, \tag{9}$$

$$-\frac{dF}{dH} = \mu_0 M = \alpha E + \mu_0\chi_v H. \tag{10}$$

For the model operating with longitudinal magnetoelectric effect (Fig. 2(a)), the charge $q$ on the surface of the medium with an area of $S$, is

$$q = DS = (\varepsilon_0 E + P)S = \varepsilon_0\varepsilon_r ES + \alpha HS \tag{11}$$

where $D = \varepsilon_0 E + P$ is the electric displacement, $\varepsilon_r = \chi_e + 1$ is the relative permittivity. The magnetic flux $\varphi$ passing through the ME medium is:

$$\varphi = BS = \mu_0(H + M)S = \alpha ES + \mu_0\mu_r HS \tag{12}$$

where B=$\mu_0(H+M)$ is the magnetic induction and $\mu_r$=$\chi_v$+1 is the relative permeability.

Under a constant or zero external $E$, a change of flux d$\varphi$ will cause a change in magnetic field $H$ by d$H = $d$\varphi/\mu_0\mu_r S$ (Eq. (12)), consequently a change in $q$ by d$q = \alpha S$d$H$ (Eq. (11)). In this case, by virtue of Eqs. (11-12), a relation between d$q$ and d$\varphi$ is established

$$dq = \frac{\alpha}{\mu_0\mu_r} d\varphi \tag{13}$$

As suggested by Mathur,[16] the two-terminal model completely fulfil the original definition for the fourth linear circuit element, promising the function to linearly convert $q$ into $\varphi$. Additionally, we should point out that, in the case of a constant or zero external $H$, a varying d$q$ will induce conversely a change of d$\varphi$ by:

$$d\varphi = \frac{\alpha}{\varepsilon_0\varepsilon_r} dq \tag{14}$$

Similarly, for the two-terminal model device that operates with transverse ME effect (Fig. 2(b)), supposing that the area of the ME medium beneath electrodes is $S$, the area of the flank side is $S'$. In this case, two direct relationships between d$q$ and d$\varphi$ are established as

$$d\varphi = \frac{S'}{S}\frac{\alpha}{\varepsilon_0\varepsilon_r} dq \tag{15}$$

and

$$dq = \frac{S}{S'}\frac{\alpha}{\mu_0\mu_r} d\varphi \tag{16}.$$

From the above deductions, it is very obvious that the two-terminal models in Figs. 2(a) and (b) completely fulfil the original definition for the fourth linear circuit element, promising the function to linearly convert charge $q$ into magnetic flux $\varphi$, and reciprocally.

Note that the ME media do not follow the Faraday's law (Eq. (2)) that $\varphi$-$q$ relationship cannot be converted to $v$-$i$ relationship in this case even though the coefficients defined in Eqs. (13-16) have the same dimension of the resistance (Ohm). Clearly, their physical meaning is definitely different from that of the resistance; they are quantities reflecting the ability of device to convert $q$ into $\varphi$ or vice versa. Another intrinsic difference between the two quantities is that the resistance is a second rank *polar* tensor while the coefficients defined in Eq. (13-16) are second rank *axial* tensors because $\alpha$ is axial tensor while $\varepsilon_r$ and $\mu_r$ are polar tensors.[21] The underlying physical origin lies in the fact that the resistance is a quantity describing a dynamic (transport) phenomena while linear magnetoelectric, dielectric and magnetic effects are static (equilibrium) phenomena. In the case of transport phenomena, due to the second law of thermodynamics, time reversal operation is **NOT** permissible, making resistance and memristance the polar tensors.[21] Unfortunately, it has caused lots of confusions that the fourth linear element defined by $q$-$\varphi$ was considered as a resistor.[1,3] Thus, it is necessary and justifiable to assign an independent name to the two-terminal linear fundamental element defined in Eqs. (9) and (10) as *transtor* (corresponding to resistor, capacitor, and inductor), and the corresponding coefficients, ($\alpha/\varepsilon_0\varepsilon_r$ and $\mu_0\mu_r/\alpha$) as ***transtance T*** (corresponding to resistance, capacitance, and inductance) to distinguish them from the memristor and memristance, respectively. We also introduced a new symbol for transtor, as shown in Fig. 1(b).

As a circuit element made from ME medium, it is easy to recognize that: (i) It is an independent passive linear and time-invariant element that is allowed in linear and time-invariant system theory,[22] like all the other three circuit elements. In the case of memristor, it is not applicable. (ii) Following the sign of $\alpha$, which can be either positive or negative in an ME medium, the transtance could be either positive or negative accordingly, a property distinctively different from resistance, capacitance, and inductance as they are always positive. Therefore, as shown in Figs. 3(a)-(c), the linear relationships between two basic variables to define resistance, capacitance, and inductance always have a positive slope. In contrast, there are two linear relationships with either positive or negative slopes between $q$ and $\varphi$ to deduce the value of $T$ (Fig. 3(d)). It is also a natural result of broken of the time reversal symmetry in $T$, not in the $R$, $C$, and $L$. (iii) In an ME medium, it is well known that $\alpha^2 \leq \varepsilon_0\varepsilon_r\mu_0\mu_r$ or $|\alpha/\varepsilon_0\varepsilon_r| \leq |\mu_0\mu_r/\alpha|$.[23] This means that the transtor is completely reversible only when $\alpha^2=\varepsilon_0\varepsilon_r\mu_0\mu_r$.

To make use of the transtor in an electric circuit, it can be coupled with an inductor via a circular magnetic core with infinite permeability $\mu$ to form a well-known four terminal electric network element, gyrator, as shown in Fig. 2(c).[24] As clearly pointed out by Tellegen[24] and demonstrated by J. Zhai *et al*.,[25] gyrator is also a linear, passive network element which has peculiar anti-reciprocal relationships: $v_1 = -gi_2$, $v_2 = gi_1$, where $g$ is a constant gyrator coefficient (in ohms), $v_1$, $i_1$ and $v_2$, $i_2$ are the voltage and current from the transtor and inductor, respectively. It also proves that transtor is an independent circuit element which cannot be replaced by a network composed of other three linear two-terminal elements. Notably, however, the ME component in the gyrator has never been discussed in the context of the fourth linear circuit element. Moreover, we propose that transtor can also be similarly coupled with another transtor to form a new type of four-terminal electric network element, as shown in Fig. 2(d). In this case,

$$i_2=(\alpha_1/\varepsilon_0\varepsilon_{r1})/(\mu_0\mu_{r2}/\alpha_2)i_1 =T_1/T_2 i_1 \tag{17}$$

$$i_1=(\alpha_2/\varepsilon_0\varepsilon_{r2})/(\mu_0\mu_{r1}/\alpha_1)i_2=T_2/T_1 i_2 \tag{18}$$

where $i_1$, $T_1$ and $i_2$, $T_2$ are the current, transtance from the first and second transtor, respectively. Therefore, it can be regarded as a current converter.

## 3. Theoretical prediction of the fourth memory circuit element

In analogy to the memory counterparts of other fundamental circuit elements, namely, the memcapacitor ($M_C$) and meminductor ($M_L$), the memristor should be treated as the memory counterpart of resistor (Fig. 1(b)), to be consistent with the convention. Thus, it is appropriate to adopt $M_R$ to replace the original symbol $M$. In the place of $M$ in Fig. 1(a), we introduced the symbol $T$ to linearly relate $\varphi$ and $q$. The memory counterpart of transtor, dubbed **memtranstor**, is denoted as **$M_T$**.

Very similar to the memristive system,[2,15] a charge-driven memtranstor can be defined by:

$$\varphi(t) = M_T(x, q, t)q(t) \tag{19}$$

$$dx/dt = f(x, q, t), \tag{20}$$

where $x$ can be a set of state variables and $f$ is a continuous vector function. In a simplified case, Eqs. (19) and (20) are reduced to:

$$\varphi(t) = M_T(\int_{t_0}^{t} q(\tau)d\tau)q(t). \tag{21}$$

Similarly, a flux-driven memtranstor is defined by:

$$q(t) = \varphi(t)/M_T(\int_{t_0}^{t} \varphi(\tau)d\tau). \tag{22}$$

The typical behavior of the memristor, described by similar equations, is characterized by a "pinched hysteretic loop" dwelling in the I and III quadrants due to the uniquely positive value of $M_R$ (Fig. 3(e)). This is also true for memcapacitor and meminductor (Figs. 3(f) and 3(g)). It is reasonable to expect that, for a periodic charge/flux input, the memtranstor should also show a pinched hysteretic loop. However, $M_T$ will certainly manifest a distinctive memory behaviour as transtance can be both positive and negative (Fig. 3(d)). Consequently, the memory behaviors of a memtranstor should be more interesting. As we postulate in Fig. 3(h), the pinched hysteretic loop can lie either in the I and III quadrants or in the II and IV quadrants. Moreover, we propose that a unique butterfly-shaped pinched loop is anticipated by virtue of changed sign of $M_T$ when a sufficiently large input of charge or flux is supplied, see Fig. 3(h). The memtranstor can also be used in the gyrator and the current converter shown in Figs. 2(c) and 2(d) by replacing the transtors to have versatile memory functionalities.

## 4. Construction of the complete relational graph and perspectives

In summary, the true fourth fundamental linear circuit elements, transtor ($T$), and its memory element, memtranstor ($M_T$), can be realized by employing the ME effects. With these newly defined elements, we are able to construct a complete and harmonized fundamental circuit relational graph, which consists of four linear ($R$, $C$, $L$, $T$) and four memory ($M_R$, $M_C$, $M_L$, $M_T$) elements, as shown in Fig. 1(b). The complete relational graph for fundamental circuit elements can provide a new guide to circuit functionality. As the transtance and the memtranstance can be both positive and negative, this will largely extend the possibility of their implementation. For hundreds years, the circuit functionalities have been dominated by the regime of three linear elements. The memory elements such as memresistor are attracting more attention because they enable more advanced circuit functions. In the future, the rest of this relational graph should be fully exploited and utilized to broaden circuit functionality for next-generation intelligent devices.

**Acknowledgments**

We thank S. H Chun and K. H. Kim for their help in discussion and figure painting.

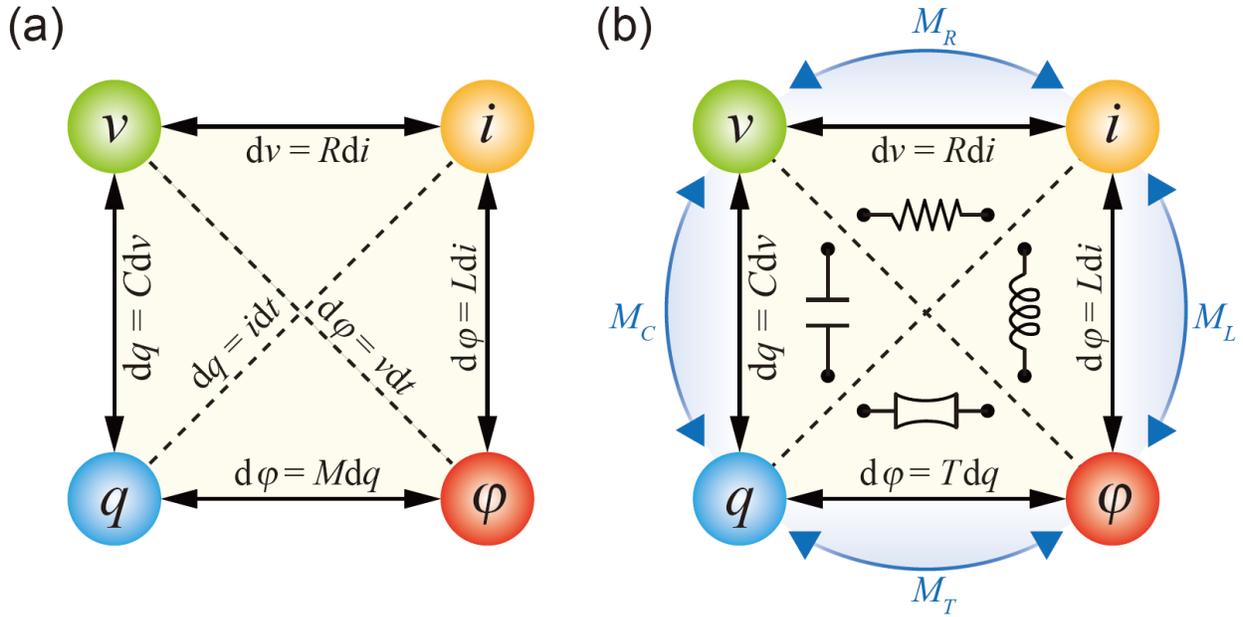

**Fig. 1**. The complete relational graph of fundamental two-terminal circuit elements. (a) The four linear fundamental two-terminal circuit elements which correlate a particular pair of the four basic circuit variables, *i.e.*, the charge $q$, the voltage $v$, the current $i$, and the magnetic flux $\varphi$. (b) A complete relational graph of all the possible fundamental two-terminal circuit elements, both linear and nonlinear. In (b), the missing fourth element, the transtor, which is anticipated to directly correlate $\varphi$ with $q$ is now denoted with $T$ in place of $M$ in (a). A symbol for transtor is introduced to facilitate later usage. The nonlinear memory devices corresponding to the four linear fundamental elements, *i.e.*, the resistor ($R$), the capacitor ($C$), the inductor ($L$), and transtor ($T$), are now accordingly termed as memristor ($M_R$), memcapacitor ($M_C$), meminductor ($M_L$), and memtranstor ($M_T$). The relations of $d\varphi = vdt$ and $dq = idt$ in (a) have been removed in (b) because they are unnecessary to define the fundamental circuit elements.

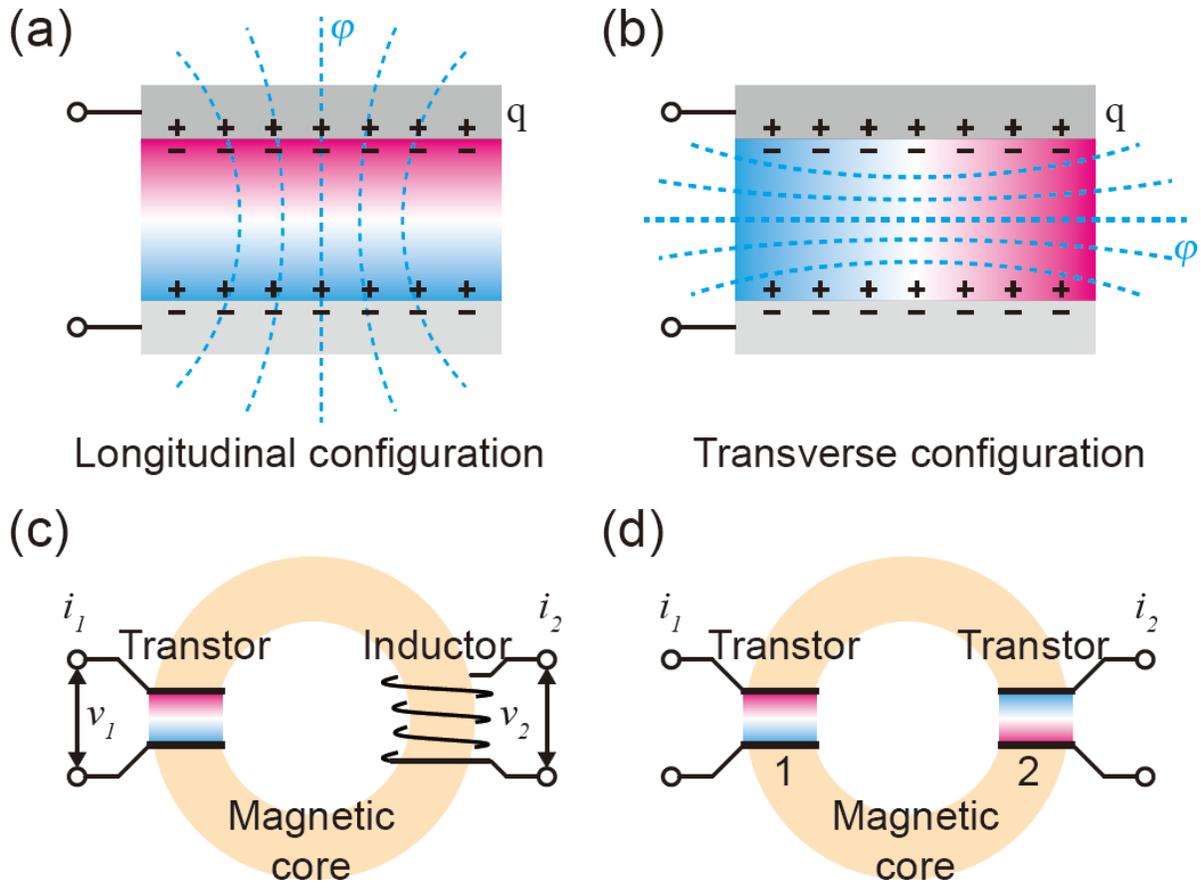

**Fig. 2.** Schematic illustration of the transtive device and its exemplar implementation. (a)-(b), Model transtive devices comprising a ME medium sandwiched between two parallel electrodes, realized with either a longitudinal (a) or a transverse (b) ME effect. (c) A four-terminal device design, the gyrator, with a transtor in the longitudinal configuration coupled to an inductor. (d) A four-terminal current converter design with a transtor in the longitudinal configuration coupled to another transtor. The memtranstor can also be used in the gyrator and the current converter by replacing the transtors to have versatile memory functionalities.

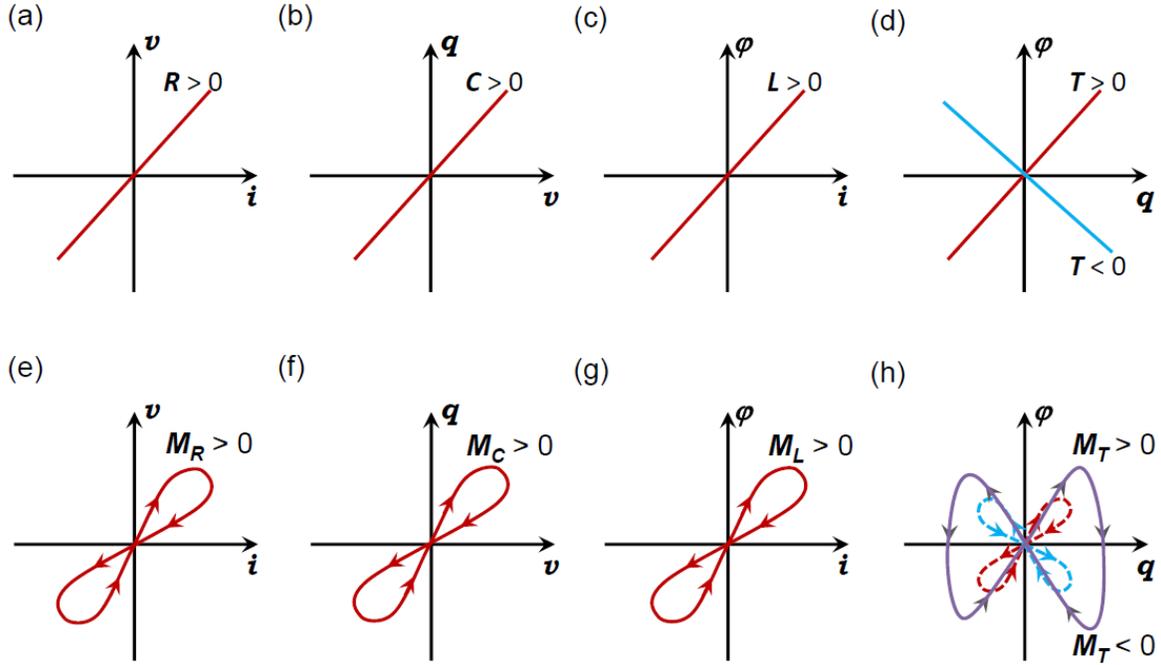

**Fig. 3.** Schematic illustration of the linear and memristive-like behaviours for all the four fundamental circuit elements. The linear relationships of the four basic circuit variables define (a), the resistance, $R$, (b), the capacitance, $C$, (c), the inductance, $L$, and (d), the transtance, $T$, with only the transtance $T$ being both positive and negative. Accordingly, (e), the memristance, $M_R$, (f), the memcapacitance, $M_C$, (g), meminductance, $M_L$, and (h), the memtranstance, $M_T$, are defined by the pinched loops of the basic circuit variables. For $M_T$, the butterfly-shaped loop in (h) occurs at a sufficiently large input of $q$.